\documentclass[%
 reprint,
superscriptaddress,
 amsmath,amssymb,
 aps,
]{revtex4-2}

\usepackage{bbm}
\usepackage{graphicx}
\usepackage{dcolumn}
\usepackage{bm}
\usepackage{floatrow}
\usepackage{braket}
\usepackage[normalem]{ulem}
\graphicspath{ {./image/} }
\usepackage{hyperref}
\usepackage{xcolor}

\begin{document}

\title{SU($N$) magnetism with ultracold molecules}
\author{Bijit Mukherjee}
\email{bijit.9791@gmail.com}
\affiliation{Joint Quantum Centre (JQC) Durham-Newcastle, Department of Chemistry, Durham University, South Road, Durham, DH1 3LE, United Kingdom.}
\author{Jeremy M. Hutson}
\email{j.m.hutson@durham.ac.uk} \affiliation{Joint Quantum Centre (JQC)
Durham-Newcastle, Department of Chemistry, Durham University, South Road,
Durham, DH1 3LE, United Kingdom.}
\author{Kaden R. A. Hazzard}
\email{kaden.hazzard@gmail.com}
\affiliation{Department of Physics and Astronomy, Rice University, Houston, Texas 77005, USA}
\affiliation{Smalley-Curl Institute, Rice University, Houston, Texas 77005, USA}

\date{\today}

\begin{abstract}
Quantum systems with SU($N$) symmetry are paradigmatic settings for quantum many-body physics. They have been studied for the insights they provide into complex materials and their ability to stabilize exotic ground states. Ultracold alkaline-earth atoms were predicted to exhibit SU($N$) symmetry for $N=2I+1=1,2,\ldots,10$, where $I$ is the nuclear spin. Subsequent experiments have revealed rich many-body physics. However, alkaline-earth atoms realize this symmetry only for fermions with repulsive interactions. In this paper, we predict that ultracold molecules shielded from destructive collisions with static electric fields or microwaves exhibit SU($N$) symmetry, which holds because deviations of the s-wave scattering length from the spin-free values are only about 3\% for CaF with static-field shielding and are estimated to be even smaller for bialkali molecules. They open the door to $N$ as large as $32$ for bosons and $36$ for fermions. They offer important features unachievable with atoms, including bosonic systems and attractive interactions. 
\end{abstract}

\maketitle
 
\section{Introduction}

Quantum systems with SU($N$) symmetry offer fascinating settings for quantum many-body physics. They have been studied for the insights they provide into complex materials and their ability to stabilize exotic ground states. In practical terms, SU($N$) symmetry can be realized for particles that have $N$ quantum states if the interactions are the same for all combinations of the states.
  
Ultracold alkaline-earth atoms with nuclear spin $I$ have been predicted and observed \cite{gorshkov2010two,cazalilla2009ultracold, wu2003exact,cazalilla2014ultracold} to exhibit SU($N$) symmetry based on their nuclear spin states, with $N$ up to $2I+1$. This arises because the nuclear spins are very weakly coupled to other degrees of freedom. Experiments on these systems have
allowed the study of a wealth of phenomena: bosonization of high-spin fermions by measurements of collective modes~\cite{song2020evidence,he2020collective,pagano2014one}, flavor-selective Mott transitions in a lattice~\cite{tusi2022flavour}, and the SU($N$) Fermi-Hubbard model's equations of state~\cite{taie20126,hofrichter2016direct,pasqualetti2023equation} and short-ranged magnetic correlations~\cite{taie2022observation}. 
The symmetry is also responsible for the temperature $T=1\,\text{nK}$ reached in Ref.~\cite{taie2022observation}; this is the lowest temperature ever achieved for fermions. Unlike ordinary SU($2$) spins, quantum fluctuations need not become classical for large spin. Consequently, exotic behavior is predicted to occur in lattices as the temperature is lowered further. Predicted phases abound for different lattice geometries and $N$, including chiral spin liquids, a topological phase of matter never before observed~\cite{hermele2009mott,chen2024multiflavor}. The itinerant or doped SU($N$) Hubbard models are little explored and likely to show extremely rich phenomena. 

However, alkaline-earth-atom realizations of SU($N$) physics have important limitations. One constraint is that, to have $I\ne 0$ ($N>1$), such atoms must be fermions to satisfy the ``even-even" rule of nuclear physics~\cite{heyde2020basic}.
Another constraint is that, empirically, all the interactions in the experimentally viable ultracold alkaline-earth atoms, Sr and Yb, are repulsive, i.e., have a positive scattering length $a$~\cite{cazalilla2014ultracold}. Moreover, the important tool of magnetic Feshbach resonances that is used to tune interactions in alkali atoms is absent for ground-state alkaline-earth atoms, since they lack unpaired electrons.

Rapidly advancing experiments with ultracold molecules offer exciting possibilities for many-body physics, quantum technologies, precision measurement, and studying chemical reactions~\cite{langen2024quantum,
koch2019quantum,quemener2012ultracold,
wall:quantum_2015,cornish:quantum_2024}. A wide variety of ultracold molecules have been produced, including bialkali molecules (hereafter alkali dimers) produced by assembly of ultracold atoms, and other species produced by direct laser cooling. 

\begin{figure*}[tbp]
\includegraphics[width=0.9\textwidth]{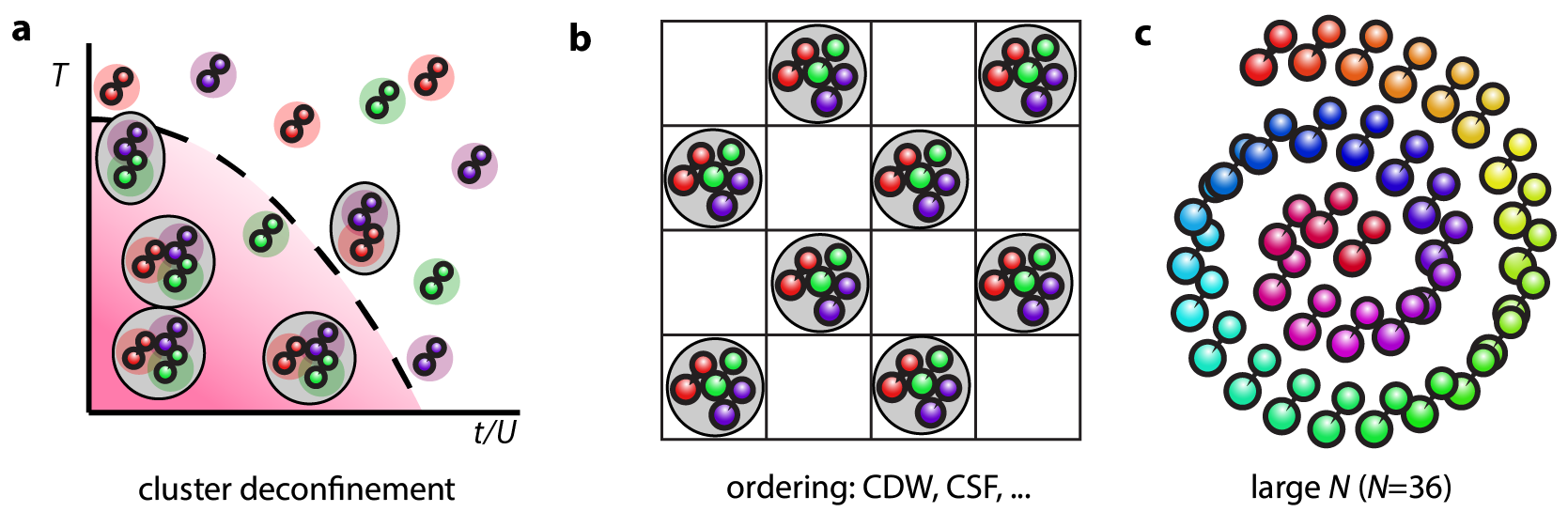}
\caption{
Applications of SU($N$) symmetry with shielded ultracold molecules: (a) formation of bound molecular clusters and their dissociation by thermal (temperature $T$) or quantum (e.g.\ lattice tunneling $t$) fluctuations, reminiscent of deconfinement in QCD, (b) ordered phases in optical lattices including charge-density waves (CDW) and color superfluids (CSF), and (c) states with large quantum fluctuations arising from large $N$ and the high degree of symmetry. 
}
\label{fig:apps}
\end{figure*}

Initial experiments on collisions of ultracold molecules revealed collisional losses that occur whenever pairs of colliding molecules reach short range ($R\lesssim 100\ a_0$). These losses occur for both reactive and non-reactive molecules and impede the creation of interesting many-body states. Their discovery stimulated theoretical proposals to ``shield" the molecules by creating intermediate-range repulsive interactions using static electric fields \cite{Avdeenkov:2006, Wang:dipolar:2015, Gonzalez-Martinez:adim:2017, Mukherjee:CaF:2023, Mukherjee:alkali:2024} or microwave radiation \cite{Karman:shielding:2018, Lassabliere:2018}. The shielding methods prevent pairs of molecules reaching short range, where most losses occur, and also suppress most inelastic collisions. Both static-field and microwave shielding have now been demonstrated experimentally \cite{Matsuda:2020, Valtolina2020, Li:KRb-shield-3D:2021, Anderegg:2021, Schindewolf:NaK-degen:2022, bigagli2023collisionally, Lin:NaRb:2023}, and two-color microwave shielding  has recently been used to achieve Bose-Einstein condensation for NaCs \cite{bigagli:observation_2023}.

In this paper, we show theoretically that shielded molecules can realize SU($N$) systems that circumvent the constraints of alkaline-earth atoms. This arises because molecules can have many spin states, and the scattering lengths for s-wave collisions of shielded molecules depend only weakly on their spin states. Fermionic and bosonic molecules are available, and the sign and magnitude of the scattering length can be tuned by varying the control fields that generate the shielding \cite{Lassabliere:2018, Mukherjee:alkali:2024}. The dipolar interactions may also be tuned.
Experimentally available molecules can realize all $N$ up to $N=36$, for Na$^{40}$K~\cite{Schindewolf:NaK-degen:2022}, much larger than for Sr ($N=10$) and Yb ($N=6$).

The properties of SU($N$) molecular systems open interesting paths for many-body physics. Attractively interacting SU($N$) systems are predicted to have rich pairing structures, for example transitions between color superfluid and trion phases in SU(3) systems~\cite{pohlmann:trion_2013, 
rapp:color_2007, 
inaba:finite_2009,
titvinidze2011magnetism,
koga2017spontaneously,
xu:trion_2023}, as illustrated in Fig.~\ref{fig:apps}(a). This has similarities to the crossover from a confining hadronic phase to a parton gas in high-energy physics~\cite{aoki2006order}. In a lattice, these trions can order, for example forming a ``charge"-density wave, as shown in Fig.~\ref{fig:apps}(b). Bosonic SU($N$) systems have garnered interest as integrable systems~ 
\cite{maassarani1998exact},  holographic duals~\cite{fujita2019effective}, and ferromagnets with non-Abelian symmetry-breaking~\cite{polychronakos2023ferromagnetic}. In all cases the large, experimentally tunable $N$ (Fig.~\ref{fig:apps}(c)) provides a control parameter, with different interesting physics potentially arising for each $N$.
Such examples offer a small glimpse of the possibilities offered by shielded ultracold molecules.

Section~\ref{sec:shielding} presents coupled-channel calculations for collisions of CaF molecules in different spin states, shielded with a static electric field. CaF is chosen because it has only 4 spin states, which makes the calculations tractable. Even CaF, which has much larger spin couplings than alkali dimers, satisfies SU($N$) symmetry to about 3\% relative accuracy, as measured by the deviations of scattering lengths for particular spin states from the spin-free value. We use the coupled-channel results to develop a model of the spin dependence that can give quantitative estimates for other molecules. The model shows that SU($N$) symmetry will hold to even higher accuracy for alkali dimers. 
Section \ref{sec:many} derives many-body models for these systems, especially the SU($N$) Hubbard models that describe shielded molecules in an optical lattice. Section~\ref{sec:outlook} summarizes, suggests experiments to verify the predictions, and outlines next steps for the field.

\section{Shielded interactions of molecules}
\label{sec:shielding}
A pair of polar molecules, $k=1,2$, interact at long range via the dipole-dipole interaction 
\begin{equation}
\hat{H}_\textrm{dd} = -\frac{3(\boldsymbol{\mu}_1\cdot\hat{\boldsymbol{R}}) (\boldsymbol{\mu}_2\cdot\hat{\boldsymbol{R}}) - \boldsymbol{\mu}_1\cdot\boldsymbol{\mu}_2}{4\pi\epsilon_0 R^3},
\label{eq:Hdd}
\end{equation}
where $R$ is the intermolecular distance, $\hat{\boldsymbol{R}}$ is the corresponding unit vector, and $\boldsymbol{\mu}_k$ is the dipole moment of molecule $k$ which lies along the molecular axis.
For an s-wave collision, with relative angular momentum (partial-wave) quantum number $L=0$, the dipole-dipole interaction averages to zero. However, it has matrix elements between $L=0$ and 2, both diagonal and off-diagonal in molecular pair state. The matrix elements off-diagonal only in $L$ cause an effective long-range attraction proportional to $d^4/R^4$ \cite{Yi:2000}, where $d$ is the space-fixed dipole moment of each molecule (for static-field shielding) or the rotating dipole (for microwave shielding).

Both static-field and microwave shielding operate by engineering a field-dressed pair state to be a small energy $\Delta E$ below the initial state of interest. 
Matrix elements of $\hat{H}_\textrm{dd}$ that connect the two pair states produce a repulsive contribution to the interaction potential for the upper (initial) state. Shielding occurs when the repulsion is sufficient to prevent pairs of colliding molecules coming close together. The repulsion is proportional to $(d^4/\Delta E)/R^6$ at long range. 

The combination of long-range attraction and shorter-range repulsion produces a potential well at long range, whose depth and position depend on the molecule and the field applied. This allows considerable control over the scattering length, which for some molecules may be tuned from positive to negative values and even through poles \cite{Lassabliere:2018, Mukherjee:alkali:2024}.

We focus here on static-field shielding, using coupled-channel scattering calculations. We begin with spin-free calculations on bosonic CaF in its vibronic ground state (X$^2\Sigma^+$, $v=0$) in Sec.\ \ref{sec:spin-free} and then consider the effects of electron and nuclear spins in Sec.\ \ref{sec:spin-inc}. We use these results to develop a semiclassical model of the effects in Sec.\ \ref{sec:model} and apply the model to experimentally available alkali dimers in Sec.\ \ref{sec:alkali}.

\subsection{The spin-free case}
\label{sec:spin-free}

We consider collisions between molecules in the state $(\tilde{n},m_n)=(1,0)$, for which static-field shielding is most effective. The single-molecule eigenstates are labeled by hindered-rotor quantum numbers ${\tilde n}$, which correlate at zero field with the free-rotor quantum number $n$, and $m_n$, its projection onto $z$. 

\begin{figure}[tbp]
\includegraphics[width=0.95\textwidth]{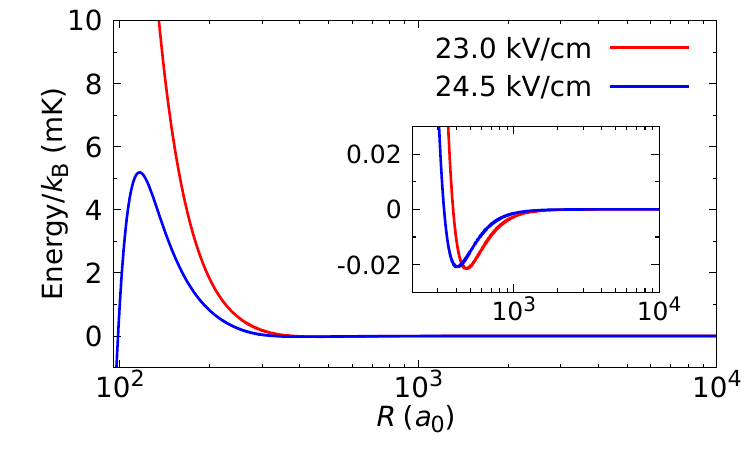}
\caption{
Effective potentials (adiabats) for spin-free CaF at the threshold $(\tilde{n},m_n) = (1,0)+(1,0)$ for fields of 23 and 24.5 kV/cm. The differences between the adiabats at different fields allow tuning of the scattering length. The inset shows an expanded view of the long-range potential well.}  
\label{fig:CaF_spin-free-adiabats}
\end{figure}

It is helpful to consider the effective potentials for scattering. We define these as adiabats that are the $R$-dependent eigenvalues $U_i(R)$ of the pair Hamiltonian given in Appendix \ref{app:spin-free}. An electric field slightly exceeding a critical value brings the pair state $(\tilde{n}_1,m_{n,1})+(\tilde{n}_2,m_{n,2})$ = (1,0)+(1,0) close to, and above, the state (0,0)+(2,0) \cite{Avdeenkov:2006}. These two pair states mix via $\hat{H}_\textrm{dd}$ to produce a repulsive interaction for (1,0)+(1,0).
Figure \ref{fig:CaF_spin-free-adiabats} shows the spin-free adiabats for s-wave scattering for two CaF molecules in state $(\tilde{n},m_n)=(1,0)$ at two values of an applied electric field $F$ that produce effective shielding. It is the variation of the adiabats with electric field that allows tuning of the scattering length.

\begin{figure}[tbp]
\includegraphics[width=\columnwidth]{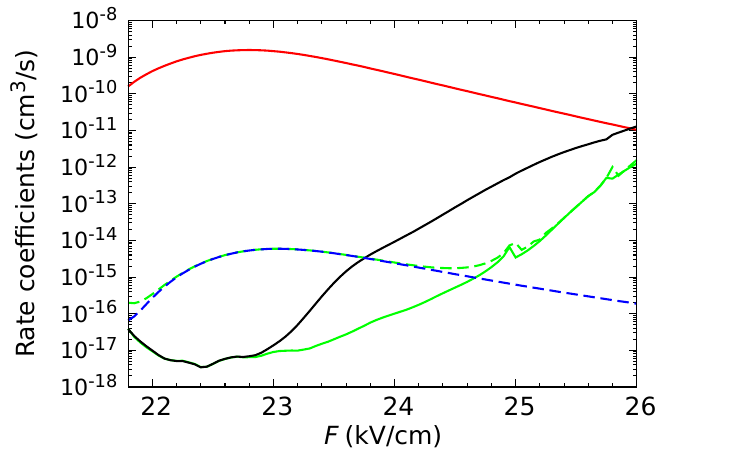}
 \caption{Collision rate coefficients for CaF. Solid lines show spin-free rate coefficients for elastic scattering (red), inelastic scattering (green) and short-range loss (black) at collision energy $E_\textrm{coll} = 10$ nK${}\times k_\textrm{B}$, over the range of electric fields $F$ where shielding is effective. The corresponding results including spin, for molecules in the initial pair state $(g,m_g)=(1,1)+(1,1)$, are shown as dashed lines. The resulting curves for elastic scattering and short-range loss lie underneath the spin-free ones. The dashed blue line shows the contribution from the 1-molecule inelastic transition to (1,0)+(1,1). The calculations including spin use the spin-N206-L6 basis set described in Appendix \ref{app:cc}.}%
    \label{fig:shield}
\end{figure}

There are two sources of 2-body loss. First, the repulsion does not extend all the way to $R=0$. There is a repulsive barrier, and some colliding pairs may tunnel through it to reach short range $(R\lesssim 100\ a_0$), where loss may occur. Secondly, there can be inelastic collisions that produce molecules in lower field-dressed states, particularly in the pair state (0,0)+(2,0), which is just below the initial one. The solid lines in Fig.\ \ref{fig:shield} show the calculated rate coefficients for spin-free elastic scattering and total (inelastic + short-range) loss for CaF as a function of $F$ \cite{Mukherjee:CaF:2023}. The overall effectiveness of shielding may be characterized by the ratio $\gamma$ of rate coefficients for elastic scattering and total loss; this can be up to $10^7$ for CaF and much larger for molecules such as NaCs \cite{Mukherjee:alkali:2024}. The calculations also provide the complex s-wave scattering length, $a(F)= \alpha(F) - i\beta(F)$, where $\beta$ arises due to loss and $L$-changing collisions and is small when shielding is effective. The real part $\alpha(F)$ is of principal interest here. Figure \ref{CaF_diff-a}(a) shows $\alpha(F)$ for CaF from spin-free coupled-channel calculations.

\subsection{Dependence on spin state}
\label{sec:spin-inc}
SU($N$) symmetry exists when 
the shielded interactions are almost independent of molecular spin state and diagonal in it. Here we describe an approach to estimate the dependence on spin state and present quantum scattering calculations for CaF that validate the estimate. We then give estimates of the spin dependence for a variety of ultracold molecules of current experimental interest.
 
Electron and nuclear spins are described by the Hamiltonian for fine and hyperfine structure, $\hat{h}_{\mathrm{fhf}}$, given in Appendix \ref{app:spin}. For CaF in a strong electric field, the electron spin $s=1/2$ and the nuclear spin $i=1/2$ of $^{19}$F couple to form $g$, with projection $m_g$; these are approximately conserved and can take values $(g,m_g)=(0,0)$, (1,0) and $(1,\pm1)$.

We first diagonalize the single-molecule Hamiltonian, including $\hat{h}_{\mathrm{fhf}}$. We calculate the space-fixed dipole moments $d_j=\langle j|\boldsymbol{\mu}_z|j\rangle$ for $j=(\tilde{n},m_n,g,m_g)$ with $(\tilde{n},m_n)=(1,0)$.
It is convenient to define fractional changes $\Delta d_j=(d_j-d_0)/d_0$ from the spin-free value $d_0$. These depend only weakly on $F$: for CaF at $F=23$ kV/cm, the values are $-1.9\times10^{-5}$, $8.3\times10^{-4}$ and $-4.4\times10^{-4}$ for the states $(g,m_g)=(0,0)$, $(1,0)$ and $(1,\pm1)$, respectively.
The coefficient of the long-range attraction is proportional to $d_j^4$ and thus differs
by at most 0.33\% from $d_0^4$. We may expect that the effective potentials will differ by about this amount.

\begin{figure}[tbp]
\includegraphics[width=\columnwidth]{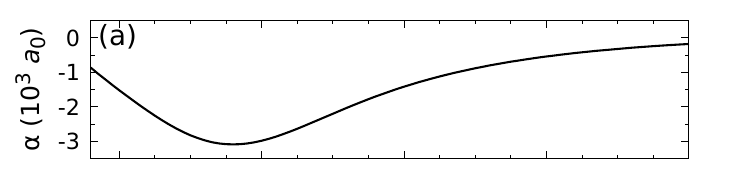}
\includegraphics[width=\columnwidth]{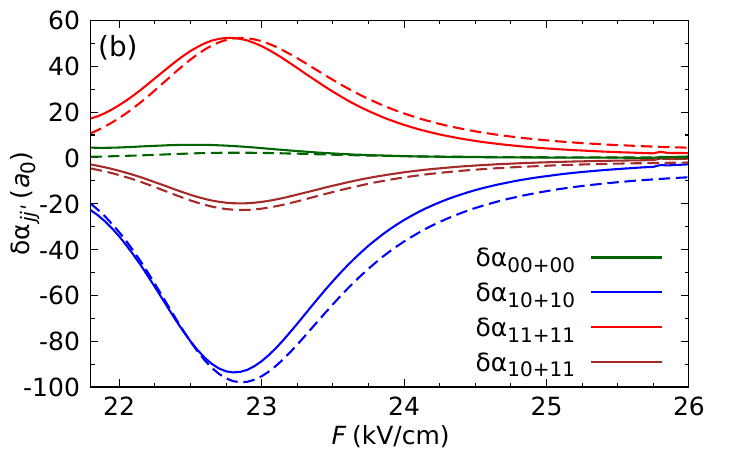}
    \caption{Effect of spins on scattering length. (a) Real part $\alpha(F)$ of the scattering length for CaF from spin-free calculations. (b) Scattering lengths $\alpha_{jj'}(F)$, including the effects of spin, shown as differences from spin-free values. Solid lines are from coupled-channel calculations; dashed lines are from the model of Eq.\ \ref{eq:dabydD-full}. Each spin combination is labeled by $gm_g+g'm'_g$.}%
    \label{CaF_diff-a}
\end{figure}

To test this simple model of the interactions, we perform full coupled-channel calculations including electron and nuclear spins for both molecules. Details are in Appendices \ref{app:cc} and \ref{app:spin}. The calculations produce scattering lengths $a_{jj'}$ and rate coefficients for elastic scattering and loss, as before, but now for each pair of spin states $j=(g,m_g)$ and $j'=(g',m'_g)$. In zero magnetic field there are 7 distinct pairs, because pairs with $(m_g,m'_g)=(0,\pm1)$ are equivalent, as are the pairs $(\pm1,\pm1)$, though the latter are different from $(\pm1,\mp1)$.

Shielding remains effective for all spin states, even for distinguishable pairs. The elastic scattering and short-range loss are almost unaffected. However, there are additional inelastic transitions for some spin states due to spin-changing collisions. In particular, a molecule initially in $(g,m_g)= (1,1)$ can undergo a transition to (1,0). The rate coefficient for this process in a collision of two such molecules is shown as a blue dashed line in Fig.\ \ref{fig:shield}. This dominates the total inelastic loss in the shielding region, though inelastic transitions to other rotor pair states dominate at $F>25$ kV/cm, where shielding is ineffective.
The spin-changing rate coefficients are no larger than $10^{-14}$ cm$^3$ s$^{-1}$ and the ratio $\gamma$ remains above $\sim{}10^5$. The spin-changing rates are similar for other collisions involving a molecule in state (1,1), and otherwise very small. This satisfies the requirement that the interactions are diagonal in spin state. 

The real part of the scattering length $\alpha(F)$ depends only weakly on spin state. The solid lines in Fig.\ \ref{CaF_diff-a}(b) show the differences $\delta \alpha_{jj'}(F)=\alpha_{jj'}(F)-\alpha_0(F)$ between the scattering lengths and the spin-free value $\alpha_0(F)$ of Fig.\ \ref{CaF_diff-a}(a) as a function of field. The values $\alpha_{jj'}$ not shown are close to $(\alpha_{jj}+\alpha_{j'j'})/2$. This demonstrates that the scattering lengths are independent of spin state to within about 5\% for shielded CaF.

\begin{figure}[tbp]
		\includegraphics[width=\columnwidth]{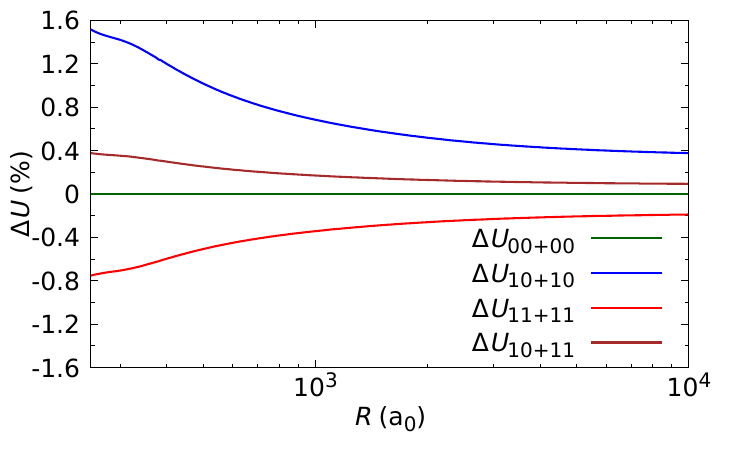}
    \caption{Effect of spins on the adiabats. Fractional differences $\Delta U_{jj'}(R)$ between adiabats with and without spin, defined by Eq.\ \ref{eq:DeltaU}, for CaF at 23 kV/cm. Each spin combination is labeled by $gm_g+g'm'_g$.}%
    \label{fig:CaF_spin-adiabats}
\end{figure}

The coupled-channel calculations provide adiabats as in Fig.\ \ref{fig:CaF_spin-free-adiabats}, but now for each spin combination. They are almost indistinguishable on the scale of Fig.\ \ref{fig:CaF_spin-free-adiabats}. However, they cross zero at slightly different inner turning points $R_{\textrm{t}jj'}$, so to show their differences we define
\begin{equation}
\Delta U_{jj'}(R) = \frac{U_{jj'}(R+R_{\textrm{t}jj'}-R_{\textrm{t}0})-U_0(R)}{U_0(R)},
\label{eq:DeltaU}
\end{equation}
 shifting $U_{jj'}(R)$ slightly in $R$ so that its turning point matches $R_{\textrm{t}0}$.
Fig.\ \ref{fig:CaF_spin-adiabats} shows $\Delta U_{jj'}(R)$ for all spin combinations of CaF at 23 kV/cm. The differences between the adiabats including spin and the spin-free adiabat are no more than 2\% over the entire classically allowed range of $R$. 
The effective potential for interaction of molecules in spin states $j$ and $j'$ has long-range form 
\begin{equation}
U_{jj'}(R) = -\frac{4\hbar^2 D_{jj'}^2}{15\mu_\textrm{red}R^4},
\end{equation}
where $D_{jj'}=d_j d_{j'} \mu_\textrm{red}/(4\pi\epsilon_0\hbar^2)$ is the dipole length for space-fixed dipoles $d_j$ and $d_{j'}$, and $\mu_\textrm{red}$ is the reduced mass.
At large $R$, the ratios of the adiabats are accurately given by the corresponding ratios of $D_{jj'}^2$.

\subsection{Effective-potential model of spin dependence}
\label{sec:model}

\begin{figure}[tbp]
\includegraphics[width=\columnwidth]{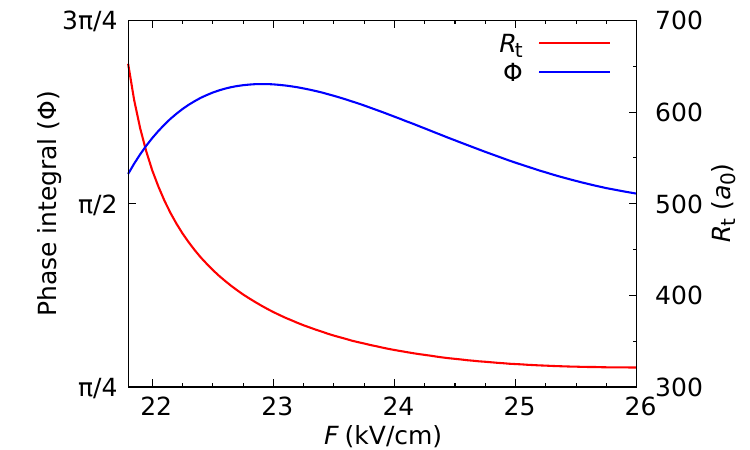}
    \caption{Features of the spin-free effective potential for CaF. The blue curve shows the phase integral $\Phi$ over the range of electric fields $F$ where shielding is effective. The red curve shows the corresponding inner turning point $R_\textrm{t}$.}%
    \label{fig:Phi-tp}
\end{figure}

In a semiclassical approximation \cite{Gribakin:1993}, the s-wave scattering length $a$ for a single channel with a long-range potential proportional to $R^{-4}$ may be written in terms of a phase integral $\Phi$,
\begin{equation}
a = R_\textrm{t} - \sqrt{8/15} D \tan\left(\Phi-\frac{\pi}{4}\right).
\label{eq:a-Phi}
\end{equation}
Here $R_\textrm{t}$ is the inner turning point at zero collision energy, with $U(R_\textrm{t})=U(\infty)=0$, $D$ is the dipole length, and
\begin{equation}
\Phi = \int_{R_\textrm{t}}^\infty k(R)\,dR,
\end{equation}
where $k(R)=(2\mu_\textrm{red}|U(R)|/\hbar^2)^{1/2}$. 
The first term in Eq.\ \ref{eq:a-Phi}, $R_\textrm{t}$, is omitted in Ref.\ \cite{Gribakin:1993}, but can be a substantial fraction of $D$ for shielded collisions; for CaF at 23 kV/cm, with $D\approx 1290~a_0$, $R_\textrm{t}$ contributes about 380 $a_0$ to the scattering length. Both $\Phi$ and $R_\textrm{t}$ are shown for CaF in Fig.\ \ref{fig:Phi-tp}, as a function of field.

If we take the spin-free adiabat $U_0(R)$ as a reference, with phase $k_0(R)$ and phase integral $\Phi_0$, the integral $\Phi_{jj'}$ for a slightly shifted potential $U_{jj'}(R)$ for the interaction between species in spin states $j$ and $j'$ is
\begin{align}
\Phi_{jj'} =& \int_{R_{\textrm{t}jj'}}^\infty k_{jj'}(R)\,dR \nonumber\\
\approx& \int_{R_\textrm{t0}}^\infty k_0(R) \left[1 + \textstyle{\frac{1}{2}} \Delta U_{jj'}(R)\right]\,dR.
\label{eq:Phi}
\end{align}

\begin{figure}[tbp]	\includegraphics[width=0.8\textwidth]{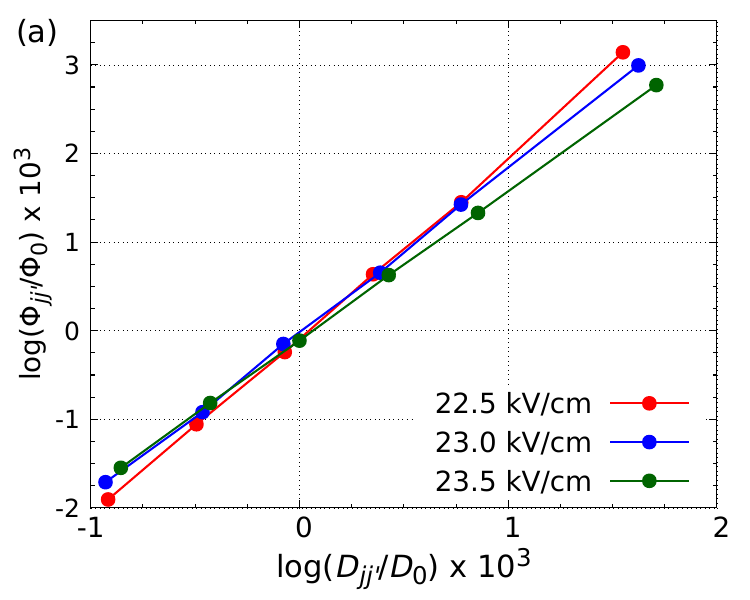}
\includegraphics[width=0.8\textwidth]{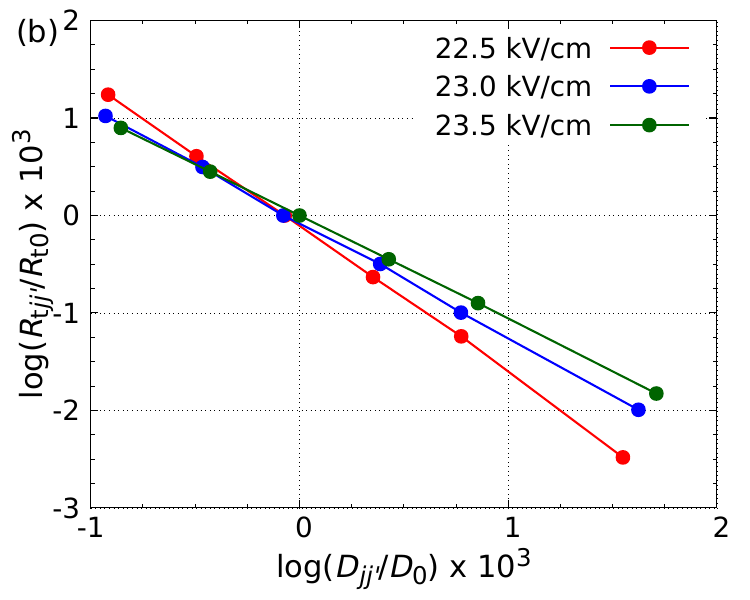}
\caption{(a) Phase integrals $\Phi_{jj'}$ and (b) inner turning points $R_{\textrm{t}jj'}$ for different spin states $j$ as a function of dipole length $D_{jj'}$, for three different electric fields in the shielding region. 
All quantities are expressed as ratios to their spin-free values.
The slopes of the plots show that, at each field, $\Phi_{jj'}$ scales approximately as $D_{jj'}^2$ and $R_{\textrm{t}jj'}$ scales approximately as $D_{jj'}^{-1}$.
\label{fig:log-log_phi_Rt}}
\end{figure}

As shown in Fig.\ \ref{fig:log-log_phi_Rt}, $\Phi_{jj'}$ scales approximately with $D_{jj'}^2$ and $R_\textrm{t}$ scales approximately with $D_{jj'}^{-1}$.
These dependencies apply to varying spin combination at constant field, but \emph{not} to varying field, which also changes the separation of the field-dressed states.

Differentiating Eq.\ \ref{eq:a-Phi} with these dependencies on $D$ gives
\begin{align}
\frac{da}{dD} \approx& -R_\textrm{t}/D - \sqrt{8/15} \tan\left(\Phi-\frac{\pi}{4}\right) \nonumber\\
&- 2 \Phi \sqrt{8/15} \sec^2\left(\Phi-\frac{\pi}{4}\right)
\nonumber\\
\approx& (a-2R_\textrm{t})/D - 2 \Phi \sqrt{8/15} \sec^2\left(\Phi-\frac{\pi}{4}\right).
\label{eq:dabydD-full}
\end{align}
The last term is large near any poles in $a$; it has minima near $\Phi/\pi=\hbox{integer} + 1/4$, which is close to the zeroes in $a$, but is nevertheless usually the largest term under shielding conditions. 

For the coupled-channel problems involved in shielding, the arguments above apply to the
real part $\alpha$ of the scattering length. The variations in $\alpha$ from its spin-free value $\alpha_0$ may be approximated
\begin{equation}
\delta\alpha_{jj'}= (D_{jj'}-D_0) \frac{d\alpha}{dD} \approx D_0 \left(\Delta d_j + \Delta d_{j'}\right) \frac{d\alpha}{dD},
\label{eq:Delta-alpha}
\end{equation}
where $D_0=d_0^2 \mu_\textrm{red}/(4\pi\epsilon_0\hbar^2)$.

\begin{table}[tbp]
\caption{Comparing changes in scattering lengths predicted by the model with coupled-channel calculations including spin. For CaF at 23 kV/cm, $D_0 = 1292~a_0$, $R_{\textrm{t}0} = 381.7~a_0$, $\Phi_0/\pi = 0.6628$ and $\alpha_0= -2980\ a_0$. All lengths are in units of $a_0$.
\label{tab:dadD}}
\centering
\begin{ruledtabular}
\begin{tabular}{ccccrr}
$(g,m_g)+(g^{\prime},m_g^{\prime})$ & $D_{jj'}$ & $R_{\textrm{t}jj'}$ & $\Phi_{jj'}/\pi$ & $\delta\alpha_{jj'}$ & $\delta\alpha_{jj'}$ \\
&  &  &  & (model) &  (c.c.) \\
\hline
(0,0)+(0,0) & 1292 & 381.7 & 0.6628 & 2.22 & 4.28 \\
(0,0)+(1,0) & 1293 & 381.4 & 0.6638 & $-$46.5 & $-$41.7 \\
(0,0)+(1,1) & 1292 & 381.9 & 0.6622 & 26.6 & 26.3 \\
(1,0)+(1,0) & 1294 & 381.0 & 0.6648 & $-$95.3 & $-$88.7 \\
(1,0)+(1,1) & 1293 & 381.6 & 0.6633 & $-$22.1 & $-$19.2 \\
(1,1)+(1,1) & 1291 & 382.1 & 0.6617 & 51.0 & 48.8 \\
(1,1)+(1,$-1$) & 1291 & 382.1 & 0.6617 & 51.0 & 48.8 \\
\end{tabular}
\end{ruledtabular}
\end{table}

The dashed lines in Fig.\ \ref{CaF_diff-a} show the results of the model for $\delta\alpha_{jj'}$, compared to the coupled-channel results including spin for CaF (solid lines). The model captures the overall behavior well. Details of the model for CaF at 23 kV/cm are given in Table \ref{tab:dadD}, including the spin combinations not shown in Fig.\ \ref{CaF_diff-a}. At this field, $\Phi$ for CaF is slightly less than $3\pi/4$; this is close to a pole in $\alpha$ as a function of $\Phi$, with $\alpha$ large and negative. Here the secant term in Eq.\ \ref{eq:dabydD-full} dominates, and small changes in $\Phi$ cause large fractional changes in $\alpha$. Nevertheless, even for CaF at 23 kV/cm, where $d_j$ varies by up to 0.13\% between spin states, the values of $\delta\alpha_{jj'}$ are no more than 3\% of $\alpha$. For systems where $\Phi$ is not close to a pole, there will be less amplification of $d\alpha/dD$ by the secant term in Eq.\ \ref{eq:dabydD-full}.

\begin{figure}[tbp]
   \includegraphics[width=\textwidth]{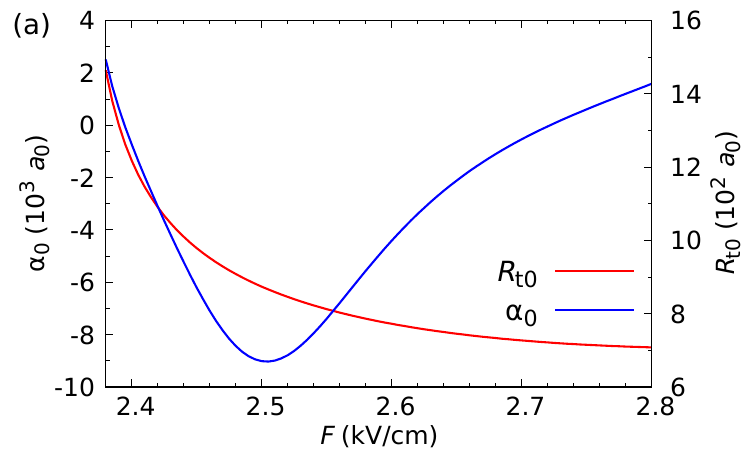}
   \includegraphics[width=\textwidth]{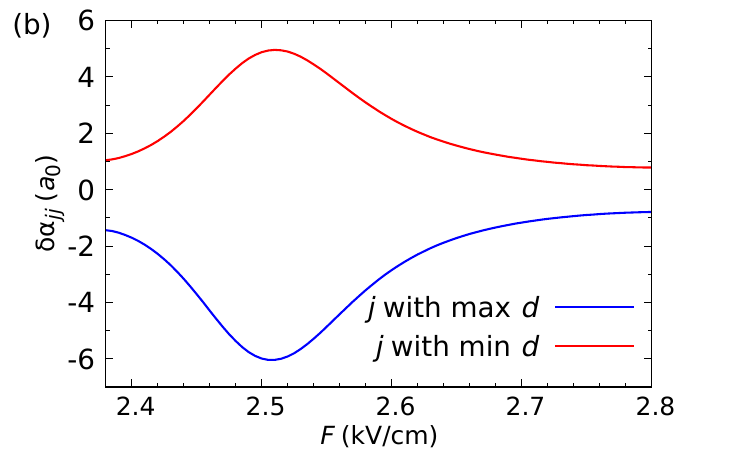}
    \caption{Predicting spin-dependent changes in scattering length for NaCs from the model. (a) Spin-free turning point $R_{\textrm{t}0}$ and real part of scattering length $\alpha_0$ for NaCs at fields where shielding is effective; (b) Real parts of $\alpha_{jj}$ from Eq.\ \ref{eq:dabydD-full}, including the effects of spin, shown as differences $\delta\alpha_{jj}$ from spin-free values. Results are shown for spin states $j$ with maximum and minimum values of $d_j$ and span the range of possible values of $\delta\alpha_{jj'}$.}%
    \label{fig:NaCs_a-model}
\end{figure}

\subsection{Extension to alkali dimers}
\label{sec:alkali}

CaF has only 4 spin states, so can be used to realize SU($N$) up to $N=4$. Other molecules, particularly alkali dimers, can reach much greater $N$. A singlet molecule AB with nuclear spins $i_\textrm{A}$ and $i_\textrm{B}$ has $(2i_\textrm{A}+1)(2i_\textrm{B}+1)$ spin states. Thus alkali dimers with two spin-3/2 nuclei, such as Na$^{87}$Rb or Na$^{39}$K, can reach $N=16$, while bosonic molecules with larger spins, such as NaCs, can reach $N=32$. Fermionic Na$^{40}$K can reach $N=36$.

\begin{table}[tbp]
\caption{Range of dipole moments $(d_\textrm{max}-d_\textrm{min})/d_0$ for different spin states of the field-dressed level $(\tilde{n},m_n)=(1,0)$ for alkali dimers and CaF. For each molecule, we choose an electric field where static-field shielding is effective, but the ranges of dipole are only weakly dependent on field. The value of $R_{\textrm{t}0}$ is also given at the chosen field, but this depends much more strongly on field.
\label{tab:Deltaj-range}}
\centering
\begin{ruledtabular}
\begin{tabular}{lccc}
Molecule & $F$ (kV/cm) & $|(d_\textrm{max}-d_\textrm{min})/d_0|$ & $R_{\textrm{t}0}$ $(a_0)$ \\
\hline
RbCs & 2.7 & $4.3 \times 10^{-4}$ & 750 \\
Na$^{39}$K & 7.1 & $8.3 \times 10^{-5}$ & 630\\
Na$^{40}$K & 7.1 & $8.9 \times 10^{-5}$ & a\\
Na$^{41}$K & 7.0 & $1.0 \times 10^{-4}$ & 620\\
NaRb & 4.5 & $4.9 \times 10^{-4}$ & 650\\
NaCs & 2.5 & $2.4 \times 10^{-5}$ & 870\\
CaF  & 23  & $1.3 \times 10^{-3}$ & 380\\
\end{tabular}
\end{ruledtabular}
$^\textrm{a}$For Na$^{40}$K, which is fermionic, the lowest channel has $L=1$ and its adiabat never drops below zero energy.
\end{table}

Alkali dimers have a hyperfine Hamiltonian \cite{Aldegunde:polar:2008} similar to that for CaF, but with the electron spin replaced by a second nuclear spin and additional terms arising from nuclear quadrupole coupling. Coupled-channel calculations that fully include spin are challenging for these molecules, because the spin space is so much larger: for NaCs, for example, the number of spin functions for the pair is 64 times larger than for CaF, and the computer time scales as the cube of this. Nevertheless, we can estimate the spin dependence, using the effective-potential model of Eq.\ \ref{eq:Delta-alpha}. 

The range of scattering lengths for different spin states depends on the range of space-fixed dipole moments $d_j$ for the molecule concerned. Table \ref{tab:Deltaj-range} summarizes the fractional changes in $d_j$ across spin states for several ultracold molecules of current interest, for the field-dressed level $(\tilde{n},m_n)=(1,0)$ that can be shielded with a static electric field. All the alkali dimers have ranges substantially smaller than CaF. 

NaCs is particularly interesting, because $\alpha$ can be tuned close to zero at $F \approx 2.395$ kV/cm \cite{Mukherjee:alkali:2024}, in the region where shielding is effective. It has 32 hyperfine states and a particularly small range of dipoles, because NaCs has unusually small nuclear quadrupole coupling constants \cite{Aldegunde:singlet:2017}. Figure \ref{fig:NaCs_a-model}(a) shows $R_\textrm{t0}(F)$ and $\alpha_0(F)$ from spin-free coupled-channel calculations on NaCs. Figure \ref{fig:NaCs_a-model}(b) shows $\delta\alpha_{jj}(F)$ from Eq.\ \ref{eq:dabydD-full} for the hyperfine states with the largest and smallest values of $d_j$. The variation in $\alpha_{jj'}$ between spin states is only about 0.1\% at most fields, and only about 3~$a_0$ around the zero in $\alpha$.

\section{Many-body physics}
\label{sec:many}
Molecular systems with an SU($N$) symmetry offer vast new possibilities for quantum simulation and many-body physics. The large spin degeneracy and the high symmetry enhance quantum fluctuations, stabilize exotic states of matter such as chiral spin liquids~\cite{hermele2009mott,chen2024multiflavor}, and produce interesting dynamics, such as controllable prethermalization~\cite{huang:suppression_2020}. Experiments will fall into two categories: experiments in continuous space with just a trap, and optical lattice experiments.

In continuum experiments, the use of molecules will enrich the SU($N$) phenomena studied with alkaline-earth atoms and also allow exploration of totally new areas. Ref.~\cite{christakis2023probing} has demonstrated quantum gas microscopy, which remains in development for fermionic alkaline-earth atoms. The large number of hyperfine states
will allow exploration of repulsive SU($N$) models with much larger $N$ than for alkaline-earth atoms; this will enhance quantum fluctuations and topological order~\cite{hermele2009mott,chen2024multiflavor}. 
The larger $N$ may also allow even lower temperatures than in alkaline-earth atoms, which already reach record-low temperatures for fermions~\cite{taie2022observation}. 

Attractive and bosonic systems are also rich areas. Attractive gases, both with and without an optical lattice, may allow experiments to explore the formation of energetically favorable clusters and their ordering, with connections to both condensed matter~\cite{rapp:color_2007, 
inaba:finite_2009, titvinidze2011magnetism,
koga2017spontaneously,
pohlmann:trion_2013,xu:trion_2023} and high-density nuclear matter~\cite{aoki2006order}. Bosonic SU($N$) systems have been considered, for example as integrable systems~\cite{maassarani1998exact}, non-Abelian ferromagnets~\cite{polychronakos2023ferromagnetic} and holographic duals~\cite{fujita2019effective}.

To explore these areas efficiently and connect experiments to models studied in many-body physics, it is necessary to reduce the coupled-channel results to an effective interaction. 
This is analogous to replacing the complicated interatomic potential for atoms with a delta-function. Due to the range of thousands of bohr and $1/R^4$ tail, a contact potential is probably adequate only for very dilute gases. 
The strength of the delta-function interaction can be determined from the scattering length in the coupled-channel calculations. Higher densities will probably require more accurate effective potentials based on the complete adiabats.
 
Optical lattice experiments with molecules provide another wide-ranging arena for SU($N$) many-body physics. In a sufficiently deep lattice, with temperatures and interactions small compared to the band gap, the system is described by an SU($N$) Hubbard Hamiltonian, 
\begin{align}\label{eq:SUN-Hubbard}
{\hat H} &= -t \sum_{\langle i,j \rangle, \sigma} \left( {\hat c}_{i \sigma}^\dagger {\hat c}_{j \sigma}^{\phantom{\dagger}} + \mathrm{h.c.} \right) + \frac{U}{2} \sum_{i,\sigma, \tau} {\hat n}_{i \sigma} {\hat n}_{i \tau} \nonumber \\
&\hspace{0.25in}{}+ \sum_{ij;\sigma\tau} \frac{V_{ij}}{2} n_{i\sigma} n_{j\tau} + \sum_{i\sigma}(\epsilon_\sigma-{\mu_\sigma}) {\hat n}_{i\sigma},
\end{align} 
where $c_{i\sigma}^{\phantom{\dagger}}$ and $c_{i\sigma}^\dagger$ are annihilation and creation operators at site $i$ for hyperfine state $\sigma$, and $n_{i\sigma}=c^{\dagger}_{i\sigma} c_{i\sigma}^{\phantom \dagger}$, $V_{ij} = C_3(1-3\cos^2(\Theta_{ij}))/|\boldsymbol{r}_i - \boldsymbol{r}_{j}|^3$ with $\Theta_{ij}$ the angle between the intermolecular separation and the electric field, is the dipole interaction between molecules $i$ and $j$, and $\epsilon_\sigma$ and $\mu_\sigma$ are the number operator, single-molecule energy and chemical potential for component $\sigma$.
The tunneling energy $t$ is
\begin{equation}
t = -\int \! d^3 r \, w^*(\boldsymbol{r}) \left( -\frac{\hbar^2}{2m}+V(\boldsymbol{r}) \right) w(\boldsymbol{r}+\boldsymbol{d}), \label{eq:Hubbard-t}
\end{equation} 
where $\boldsymbol{d}$ is a nearest-neighbor lattice vector, $m$ is the molecular mass, $V(\boldsymbol{r})$ is the lattice potential of a molecule at center-of-mass position $\boldsymbol{r}$, and $w(\boldsymbol{r})$ is the lowest-band Wannier function obtained from the single-particle band structure. 
 When the spread of the Wannier functions is much larger than the interaction range, 
\begin{equation}
U = \frac{4\pi \hbar^2 a}{m} \int \!d^3 r \, |w(\boldsymbol{r})|^4 \label{eq:Hubbard-U},
\end{equation}
where $a$ is the scattering length. 
Due to the large spatial extent of the interaction potential, Eq.\ \ref{eq:Hubbard-U} may provide only a rough estimate of $U$. Quantitative calculations of $U$ can be performed by solving the two-body problem numerically. This is a challenging calculation, but is tractable when the lattice is deep enough. 
Although the single-molecule energies in the last term of Eq.~\eqref{eq:SUN-Hubbard} apparently break SU($N$) symmetry, they are irrelevant because ${\hat N}_\sigma=\sum_i {\hat n}_{i\sigma}$, is conserved.

An important direction for future research will be to understand the phenomena that occur in molecular many-body systems with SU($N$) symmetry in optical lattices. For the particularly challenging case of repulsive fermions in optical lattices, there are a number of numerical methods that can be applied. Reference~\cite{schafer2021tracking} provides a fairly comprehensive overview and comparison of the methods for the SU(2) Fermi-Hubbard model, all of which can be extended in principle to SU($N$). Reference~\cite{ibarra2025many} reviews the numerical methods that have been applied to understand ultracold atoms with SU($N$) symmetry in optical lattices; see Table 1 therein for a summary. In parameter regimes where the long-range (off-site) dipolar interaction is important, some of these algorithms, such as determinantal quantum Monte Carlo, face serious challenges, and confronting these will be an important task for the community.

\section{Outlook \label{sec:outlook}}

Realization of SU($N$) symmetry in ultracold molecules will unlock a broad range of new physics, with strong connections to condensed matter and other areas of many-body physics. A first step will be to confirm and quantify the degree of symmetry experimentally. Initial characterizations may be performed by measuring the kinetics of evaporation or cross-dimensional thermalization, already measured for one spin component of $^{40}$KRb~\cite{Valtolina2020}. For a system with SU($N$) symmetry, these are independent of hyperfine state. Spectroscopy can provide more accurate measurements, analogous to those for SU($N$) symmetry with alkaline-earth atoms~\cite{zhang:spectroscopic_2014}. Two-photon microwave or Raman spectroscopy can measure the difference in interaction energy when the spin state is changed. A useful limit is a deep lattice where tunneling is negligible, so photons will drive one-molecule hyperfine transitions on doubly-occupied sites. Amplitude-modulation spectroscopy has also been used to measure interactions of atoms in lattices to high precision~\cite{mark:precision_2011}. With the SU($N$) symmetry confirmed, experiments can begin probing the many-body phases of matter and dynamics offered. 

This work also opens areas of theoretical research in both molecular collisions and many-body physics. Although the treatment of the alkali dimers in Sec.~\ref{sec:alkali} is sufficient to estimate the degree of SU($N$) symmetry, full coupled-channel calculations are needed for quantitative results. This challenges current methods due to the large number of nuclear states. Although we have focused here on shielding with static electric fields, we expect that microwave shielding will provide interactions with a comparable degree of SU($N$) symmetry. Coupled-channel calculations are needed to verify this.

One rich avenue offered by alkaline-earth atoms is the existence of an electronically excited ``clock state", which also exhibits SU($N$) symmetry amongst its nuclear spin states. Similar possibilities will be offered by excited vibrational states of molecules, which have very long lifetimes. The dipole moment, rotational constant and fine/hyperfine constants of molecules depend only very weakly on their vibrational state, so shielding will occur at very similar electric fields and with similar spin dependence.

The new many-body physics offered by shielded ultracold molecules includes gases, Hubbard models, and Heisenberg models with large $N$. Such systems can have positive or negative scattering lengths, bosonic or fermionic constituents, and dipolar interactions. In addition, the electric (or microwave) fields used to implement the shielding can be dynamically changed  at rates much faster than the intrinsic timescales of the  many-body  dynamics. They can also be changed much faster than is typically possible for a magnetic field, as used to control interactions via a Feshbach resonance for systems of ultracold atoms with SU(2) symmetry. This allows the possibility for experiments to study interaction quenches and ramps, and characteristic phenomena associated with them: how long-range correlations grow and spread, how short-range correlations (such as the contact) grow~\cite{Makotyn2014,sykes:quenching_2013,yin:quench_2013}, and the universal Kibble-Zurek scaling of the number of excitations created across ramps~\cite{KIBBLE1980183,zurek:cosmological_1985}. All of these new factors offer fertile ground to study new phases and dynamics of quantum matter.

\section*{Rights retention statement}

For the purpose of open access, the authors have applied a Creative Commons Attribution (CC BY) licence to any Author Accepted Manuscript version arising from this submission.

\section*{Data availability statement}

Data supporting this study are openly available from Durham University \cite{DOI_SU_N_data}.

\begin{acknowledgements}

K.R.A.H. acknowledges support from the National Science Foundation (PHY-1848304), the Robert A. Welch Foundation (C-1872), and the W. M. Keck Foundation (Grant No. 995764). K.R.A.H. benefited from discussions at the the Aspen Center for Physics, which is supported in part by the National Science Foundation (PHY-1066293). B.M. and J.M.H. are grateful to Matthew Frye and Ruth Le Sueur for valuable discussions and acknowledge support from the U.K. Engineering and Physical Sciences Research Council (EPSRC) Grant Nos.\ EP/W00299X/1, and EP/V011677/1.

\end{acknowledgements} 

\appendix

\section{Hamiltonian for the spin-free case}
\label{app:spin-free}

The Hamiltonian of spin-free CaF in an electric field is
\begin{equation}
\hat{h} = b\hat{\boldsymbol{n}}^2 - \boldsymbol{\mu} \cdot \boldsymbol{F}.
\label{eq:ham-Stark}
\end{equation}
The molecule is treated as a rigid rotor with rotational constant $b$ in an electric field $\boldsymbol{F}$ along the $z$ axis; $\hat{\boldsymbol{n}}$ is the operator for molecular rotation. 
For $^{40}$Ca$^{19}$F, $b/h \approx 10.267$\,GHz and $|\boldsymbol{\mu}|=3.07$\,D. The corresponding Hamiltonian for a pair of molecules is
\begin{equation}
\hat{h}_1 + \hat{h}_2 + \frac{\hbar^2\hat{\boldsymbol{L}}^2}{2\mu_\textrm{red}R^2} + \hat{H}_\textrm{dd} - C_6^\textrm{elec}/R^6,
\label{eq:adiabat}
\end{equation}
where $\hat{\boldsymbol{L}}$ is the operator for relative rotation of the pair and $\mu_\textrm{red}$ is the reduced mass. $C_6^\textrm{elec}$ \cite{Lepers:2013, Mukherjee:CaF:2023} accounts for the electronic dispersion interaction between the molecules, but has only small effects on the results.

\section{Basis sets used in coupled-channel calculations}
\label{app:cc}

The methodology used for the coupled-channel calculations on CaF is as described in Ref.\ \cite{Mukherjee:CaF:2023}, except that the present calculations use different basis sets.

As in Ref.\ \cite{Mukherjee:CaF:2023}, we use basis sets that are constructed from symmetrized products of field-dressed rotor functions $|\tilde{n},m_n\rangle$ and spin functions $|g,m_g\rangle$, together with functions for the partial-wave quantum number $L$ and its projection $M_L$. We include rotor functions with $\tilde{n}$ up to 5. For CaF, there are four spin functions for each monomer rotor state. 
However, the resulting number of pair basis functions, $N_\textrm{pair}$, is too large (${\sim}10000$ for each $L$, $M_L$) to be used directly in coupled-channel calculations. We therefore include only a relatively small number of “class 1” pair functions explicitly in the basis set, with the remaining “class 2” functions taken into account through Van Vleck transformations as described in ref.\ \cite{Mukherjee:CaF:2023}.
 
It is the number of class 1 functions that determines the overall computational cost. To achieve a manageable basis-set size, we include only energetically nearby rotor pairs in class 1 and move the remainder to class 2. In the present work, we 
include 14 rotor pairs in class 1: $(\tilde{n},m_n) =$ (0,0)+(1,0), (1,$-1$)+(1,$-1$), (1,1)+(1,$-1$), (1,1)+(1,1), (1,0)+(1,$-1$), (1,0)+(1,1), (1,0)+(1,0), (0,0)+(2,$-1$), (0,0)+(2,1), (0,0)+(2,0), (1,$-1$)+(2,$-1$), (1,$-1$)+(2,1), (1,1)+(2,$-1$) and (1,1)+(2,1). Inclusion of all spin functions for each of these rotor pairs gives a total number of symmetrized pair states $N_\textrm{pair}=206$. We refer to basis sets based on this as spin-N206.

The spin-dependence of scattering lengths, characterized by $\delta\alpha_{jj'}$, converges very fast with respect to the rotor basis, and is much better than 1\% for spin-N206. The convergence of loss rates is slower, and varies with field because colliding pairs are more likely to reach short range when shielding is poor. Nevertheless, at 23 kV/cm, spin-N206 gives convergence of loss rates to within 1\%.

The basis set of partial waves is also important. For each pair function (rotor plus spin), we include partial waves $L$ up to 6 and refer to the resulting basis sets as spin-N206-L6. For each spin combination, calculations are performed for only a single value of $M_\textrm{tot}=m_{n,1}+m_{n,2}+m_{g,1}+m_{g,2}+M_L$, such that the s-wave channel for the initial state is included in the basis set. The total number of coupled channels in class 1 varies from 556 to 652. Based on comparisons between spin-N206-L6 and spin-N206-L4, we estimate that spin-N206-L6 gives $\delta\alpha_{jj'}$ converged to better than 1\% and spin-changing rates converged within 10\%.
However, the spin-free loss rates are much less accurate and need $L_\textrm{max}$ as high as 20 for convergence \cite{Mukherjee:CaF:2023}. With $L_\textrm{max} = 6$, which is the highest we can afford for calculations including spins, the spin-free loss rates are underestimated by a factor of 10 at 23 kV/cm.

\section{Hamiltonian for fine and hyperfine structure}
\label{app:spin}

For a single CaF molecule, the Hamiltonian for fine and hyperfine structure is
\begin{equation}
\hat{h}_{\mathrm{fhf}} = \gamma\boldsymbol{\hat{s}} \cdot \boldsymbol{\hat{n}} + \zeta_\mathrm{F} \boldsymbol{\hat{i}} \cdot \boldsymbol{\hat{s}} + t\sqrt{6}T^2(C) \cdot T^2(\boldsymbol{\hat{i}},\boldsymbol{\hat{s}}) + c_\mathrm{F} \boldsymbol{\hat{i}} \cdot \boldsymbol{\hat{n}}.
\label{eq:ham-fhf}
\end{equation}
Here the first term represents the electron spin-rotation interaction, while the second and third terms account for the isotropic and anisotropic interactions between electron and nuclear spins. $T^2(\boldsymbol{\hat{i}},\boldsymbol{\hat{s}})$ denotes the rank-2 spherical tensor formed from $\boldsymbol{\hat{i}}$ and $\boldsymbol{\hat{s}}$, and $T^2(C)$ is a spherical tensor whose components are the Racah-normalized spherical harmonics $C^2_q(\theta,\phi)$. The last term represents the nuclear spin-rotation interaction, which is typically three orders of magnitude smaller than the others.
The values of the constants $b$, $\gamma$, $\zeta_\mathrm{F}$, $t$ and $c_\mathrm{F}$ for CaF are the same as in Ref.\ \cite{Mukherjee:CaF:2023}.

\bibliography{references,all,SU_N_data}
 
\end{document}